\documentstyle[aps,epsf,prl,multicol]{revtex}
\draft
\begin{document}
\title{Heat Transport in Turbulent Rayleigh-B\'enard Convection}

\author{Xiaochao Xu, Kapil M.S. Bajaj, and Guenter Ahlers}

\address{Department of Physics and Quantum Institute, \\University
of California, Santa Barbara, California 93106, USA}

\date{\today}\maketitle

\begin{abstract}
We present measurements of the Nusselt number $\cal N$ as a function of the Rayleigh number $R$ in cylindrical cells with aspect ratios $0.5 \leq \Gamma \equiv D/d \leq 12.8$ ($D$ is the diameter and $d$ the height). We used acetone with a Prandtl number $\sigma = 4.0$ for $10^5 \alt R \alt 4\times10^{10}$. A fit of a powerlaw ${\cal N} = {\cal N}_0 R^{\gamma_{eff}}$ over limited ranges of $R$ yielded values of $\gamma_{eff}$ from 0.275 near $R = 10^7$ to 0.300 near $R = 10^{10}$. The data are inconsistent with a single powerlaw for ${\cal N}(R)$. For $R > 10^7$ they are consistent with ${\cal N} = a \sigma^{-1/12} R^{1/4} + b \sigma^{-1/7} R^{3/7}$ as proposed by Grossmann and Lohse for $\sigma \agt 2$. 

\end{abstract}


\begin{multicols}{2}

Since the pioneering measurements by Libchaber and co-workers\cite{HCL87,CGHKLTWZZ89} of heat transport by turbulent gaseous helium heated from below, there has been a revival of interest in the nature of turbulent convection.\cite{Si94} In addition to the local properties of the flow, one of the central issues has been the global heat transport of the system, as expressed by the Nusselt number ${\cal N} = \lambda_{eff}/\lambda$. Here $\lambda_{eff} = q d / \Delta T$ is the effective thermal conductivity of the convecting fluid ($q$ is the heat-current density, $d$ the height of the sample, and $\Delta T$ the imposed temperature difference), and $\lambda$ is the conductivity of the quiescent fluid. Usually a simple powerlaw 
\begin{equation}
{\cal N} = {\cal N}_0R^{\bar \gamma}
\label{Eq:powerlaw1} 
\end{equation}
was an adequate representation of the experimental data.\cite{FN} Here  
$R = \alpha g d^3 \Delta T / \kappa \nu$
 is the Rayleigh number, $\alpha$ the thermal expansion coefficient, $g$ the gravitational acceleration, $\kappa$ the thermal diffusivity, and $\nu$ the kinematic viscosity. Various data sets yielded exponent values $\bar \gamma$ from 0.28 to 0.31.\cite{NSSD00,FN1} Most recently, measurements over the unprecedented range $10^6 \alt R \alt 10^{17}$ were made by Niemela {\it et al.}, and a fit to them of Eq.~\ref{Eq:powerlaw1} gave $\bar \gamma = 0.309$; \cite{NSSD00} but even these data did not reveal any deviation from the functional form of Eq.~\ref{Eq:powerlaw1}. Competing theoretical models involving quite different physical assumptions made predictions of powerlaw behavior with exponents $\gamma$ in the same narrow range.\cite{CGHKLTWZZ89,FN1,GL00,SS90} Here we mention just two of them. A boundary-layer scaling-theory\cite{CGHKLTWZZ89,SS90} which yielded $\gamma = 2/7 \simeq 0.2857$ was an early favorite, at least for the experimentally accessible range $R \alt 10^{12}$. It was generally consistent with most of the available experimental results. More recently, a competing model based on the decomposition of the 
kinetic and the thermal dissipation into boundary-layer and
bulk contributions was presented by Grossmann and Lohse (GL) \cite{GL00} and predicts that the data measure an average exponent $\bar \gamma$ associated with a crossover from $\gamma = 1/4$ at small $R$ to a slightly larger $\gamma$ at much larger $R$. In the experimental range the effective exponent $\gamma_{eff} \equiv d(ln({\cal N}))/d(ln(R))$, which should be compared with the experimentally determined $\bar \gamma$,  is only very weakly dependent upon $R$ and has values fairly close to 2/7. Thus, it has not been possible before to distinguish between the competing theories on the basis of the experimental data.

Here we present new measurements of ${\cal N}(R)$ over the range $10^5 \alt R \alt 4\times10^{10}$ for a Prandtl number $\sigma \equiv \nu/\kappa = 4.0$. Our data are of exceptionally high precision and accuracy. They are incompatible with the single powerlaw Eq.~\ref{Eq:powerlaw1}, and yield values of $\gamma_{eff}$ which vary from 0.277 near $R = 10^7$ to 0.300 near $R = 10^{10}$. In particular, the results rule out the prediction \cite{CGHKLTWZZ89,SS90} $\gamma = 2/7$. For $R \agt 10^7$ a good fit to our results can be obtained with the crossover function
\begin{equation}
{\cal N} = a \sigma^{-1/12} R^{1/4} + b \sigma^{-1/7} R^{3/7}
\label{Eq:powerlaw2}
\end{equation}
proposed by GL \cite{GL00} for $\sigma \agt 2$.

We used two apparatus. One was described previously\cite{DAC95}. It could accomodate cells with a height up to $d = 3$ cm. The other was similar, except that its three concentric sections were lengthened by 20 cm to allow measurements with cells as long a 23 cm. In both, the cell top was a sapphire disk of diameter 10 cm. A high-density polyethylene sidewall of circular cross section and with diameter $D$ close to 8.8 cm was sealed to the top and bottom by ethylene-propylene O-rings. Four walls, with lengths ranging from 0.70 to 17.4 cm, were used and yielded aspect ratios $\Gamma \equiv D/d = 12.8, 3.0, 1.0,$ and 0.5. The bottom plate had a mirror finish. It contained two thermistors, located 2.5 cm from the center and at 180$^\circ$ relative to each other. The fluid was acetone. Its physical properties are very well known.\cite{properties}  Measurements\cite{BA00} of the convective onset in thin cells yielded values of $R_c$ very close to 1708, thus confirming the reliability of the relevant fluid properties. Deviations from the Boussinesq approximation, which is usually assumed in theoretical work, are relatively small. From Eq. 8 of Ref. \cite{WL91} we estimated $x \simeq 1 + 0.0012\Delta T$ for the ratio of the temperature drops across the top and bottom boundary layers. On the other hand, the parameter $\cal Q$ defined by Busse \cite{Bu67}, which is usually considered near the onset of convection, can be approximated by ${\cal Q} \simeq -0.05\Delta T$ and does reach significant valued when $\Delta T$ exceeds, say 10$^\circ$C. Our temperature stability and resolution was 0.001$^\circ$C or better. We measured and corrected for the conductances of the empty cells, and applied corrections for resistance in series with the fluid. The bath temperature was controlled in a feedback loop and usually fluctuated by no more than $0.001^\circ$C rms. The bottom-plate temperature was also held constant. Its fluctuations were typically determined by the fluid turbulence, but remained less than 0.1\% of $\Delta T$. The temperatures of the two thermistors in the bottom plate could differ by up to 0.5\% of $\Delta T$ due to the large-scale flow. The parameters $\cal N$ and $R$ were calculated using the average of the two measurements. When only one was used, exponents derived from the data typically differed by no more than $4\times 10^{-4}$. Usually $\Delta T$ was stepped in equal increments on a logarithmic scale, holding the mean temperature at $32.00^\circ$C.

\narrowtext
  
\begin{figure}
\epsfxsize = 3in
\centerline{\epsffile{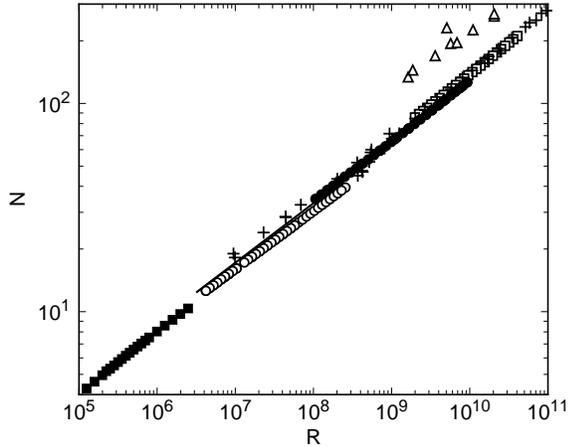}}
\caption{The Nusselt number as a function of the Rayleigh number. Solid squares: $\Gamma = 12.8$; open circles: $\Gamma = 3.0$; solid circles: $\Gamma = 1.0$; open squares: $\Gamma = 0.50$; plusses: Ref. \protect\cite{Cetal}; open triangles: Ref. \protect\cite{AS00}; solid line: Ref. \protect\cite{LE97}.}
\label{fig:N_of_R} 
\end{figure}

\begin{figure}
\epsfxsize = 3in
\centerline{\epsffile{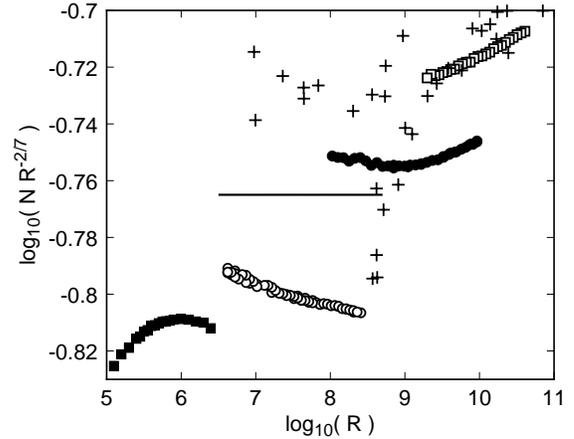}}
\caption{High-resolution plot of the Nusselt number as a function of the Rayleigh number. The symbols are as in Fig.~\ref{fig:N_of_R}.}
\label{fig:N_of_R_red} 
\end{figure}

Results of our measurements in four cells of different $\Gamma$ are shown in Fig.~\ref{fig:N_of_R}. For each cell they cover about two decades of $R$, and collectively they span the range of $R$ from $10^5$ to $4\times10^{10}$. Of interest here is the range $R \agt 10^6$, where one might expect Eq.~\ref{Eq:powerlaw1} to be relevant. There is a dependence of ${\cal N}(R)$ upon $\Gamma$, as already noted by others \cite{WL92}. For comparison, we also show in Fig.~\ref{fig:N_of_R} the recent results of Chavanne {\it et al.}\cite{Cetal} for $\sigma \simeq 0.8$ and $\Gamma = 0.5$ (plusses) and those of Ashkenazi and Steinberg \cite{AS00} for $\sigma = 1$ and a cell of square cross section and $\Gamma = 0.72$ (open triangles). There is good agreement with the former, considering the difference in $\sigma$. The latter are about a factor of 1.8 larger than our results; this difference seems too large to be attributed to the difference in $\sigma$ or the geometry and remains unexplained. The solid line just above the open circles in the figure (more easily seen in Fig.~\ref{fig:N_of_R_red}) corresponds to the fit of Eq.~\ref{Eq:powerlaw1} to the data of Liu and Ecke\cite{LE97} for $\sigma = 4$, $\Gamma \simeq 1$, and a cell with a square cross section. The agreement with our data is excellent, considering the difference in geometry.
\narrowtext

\begin{figure}
\epsfxsize = 2.75in
\centerline{\epsffile{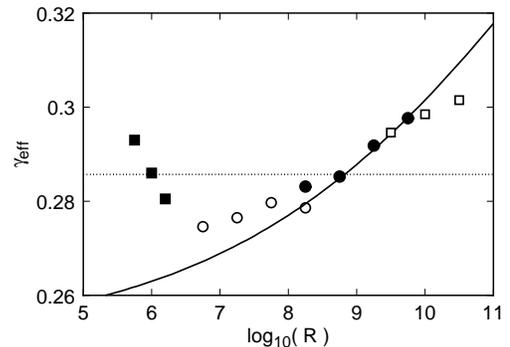}}
\caption{The effective exponent as a function of $R$. Solid squares: $\Gamma = 12.8$. Open circles: $\Gamma = 3.0$. Solid circles: $\Gamma = 1.0$. Open squares: $\Gamma = 0.5$. Solid line: logarithmic derivative of Eq.~\ref{Eq:powerlaw2} with $a = 0.326$ and $b = 2.36\times 10^{-3}$. Dotted line: $\gamma = 2/7$.}
\label{fig:gamma} 
\end{figure}

Figure~\ref{fig:N_of_R} does not have enough resolution to reveal details about the data. Thus we use the early prediction $\gamma = 2/7$ as a reference, and show $log_{10}({\cal N}R^{-2/7})$ as a function of $log_{10}(R)$ in Fig.~\ref{fig:N_of_R_red}. If the theory were correct, ${\cal N}R^{-2/7}$ should be equal to ${\cal N}_0$, {\it i.e.} independent of $R$, over its range of applicability. If Eq.~\ref{Eq:powerlaw1} is the right functional form but $\gamma$ differs from 2/7, then the data should fall on straight lines with slopes equal to $\gamma - 2/7$. Our $\Gamma = 12.8$ data (solid squares) are at relatively small $R$ and one might not expect Eq.~\ref{Eq:powerlaw1} to become applicable until $R$ is larger. The $\Gamma = 3.0$ data actually show slight curvature, but in any case would yield $\bar \gamma < 2/7$. The smaller-$\Gamma$ data are clearly curved, showing that Eq.~\ref{Eq:powerlaw1} is not applicable with any value of $\gamma$. In order to make this conclusion more quantitative, we show in Fig.~\ref{fig:gamma} effective local exponents $\gamma_{eff}$ derived by fitting Eq.~\ref{Eq:powerlaw1} to the data over various restricted ranges, each covering about half a decade of $R$. The fits yield values of $\gamma_{eff}(R)$ which have a minimum near $R = 10^7$. Interestingly, $\gamma_{eff}$ is within our resolution  independent of $\Gamma$.

\begin{figure}
\epsfxsize = 2.75in
\centerline{\epsffile{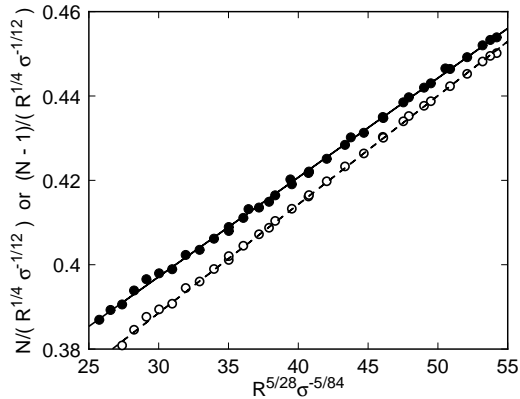}}
\caption{Plot of ${\cal N}/(R^{1/4}\sigma^{-1/12})$ (solid circles) and $({\cal N} - 1)/(R^{1/4}\sigma^{-1/12})$ (open circles) as a function of $R^{5/28}\sigma^{-5/84}$. The straight lines are least-squares fits. If Eq.~\ref{Eq:powerlaw2} is correct, the points should scatter randomly about them.}
\label{fig:GLplot} 
\end{figure}

Next we compare the predictions of GL \cite{GL00} with our data. These authors defined various scaling regimes in the $R-\sigma$ plane. For $\sigma \agt 2$, they  expect that crossover between their regions $I_u$ and $III_u$ should be observed. For that case, Eq.~\ref{Eq:powerlaw2} is predicted to apply. One way to test this \cite{Lohse} is to plot  $y = {\cal N}/(R^{1/4}\sigma^{-1/12})$ as a function of $x = R^{5/28}\sigma^{-5/84}$. If the prediction is correct, the data should fall on a straight line $y = a + bx$ with $a$ and $b$ equal to the coefficients in Eq.~\ref{Eq:powerlaw2}. Our $\Gamma = 1.0$ data are shown in this parameterization as solid circles in Fig.~\ref{fig:GLplot}. The solid line is a least-squares fit.  One sees that the data are fitted extremely well. The coefficients are $a = 0.326$ and $b = 2.36\times 10^{-3}$, in good agreement with the coefficients estimated by GL on the basis of other experimental data.
\narrowtext

We note that in the analysis it was assumed that $\cal N$ is the appropriate variable to compare with Eqs.~\ref{Eq:powerlaw1} and \ref{Eq:powerlaw2}. However, one might argue that only the convective contribution ${\cal N} - 1$ should be considered. It turns out that replacing $\cal N$ with ${\cal N} - 1$ leads to the same conclusions. The open circles in Fig.~\ref{fig:GLplot}  are obtained from our $\Gamma = 1$ data when ${\cal N} - 1$ is used. The dashed straight line is an excellent fit to the data and yields $a = 0.311$ and $b = 2.58\times 10^{-3}$. 

The logarithmic derivative $\gamma_{eff}$ of Eq.~\ref{Eq:powerlaw2} based on the analysis of $\cal N$ (rather than on ${\cal N} - 1$) is shown as a solid line in Fig.~\ref{fig:gamma}. For $R \agt 10^7$ it agrees quite well with the values determined by local fits of Eq.~\ref{Eq:powerlaw1} to the data. It would be difficult to demonstrate that the small, seemingly  systematic, deviations which do exist are outside of possible systematic errors in the measurements. For smaller $R$, one does not expect the theory to be applicable. GL estimate that their region $I_u$ has a lower boundary below which the Reynolds number $R_e \simeq 0.039R^{1/2}\sigma^{-5/6}  \alt 50$. This occurs when $R\simeq 1.6\times 10^7$.

\begin{figure}
\epsfxsize = 2.75in
\centerline{\epsffile{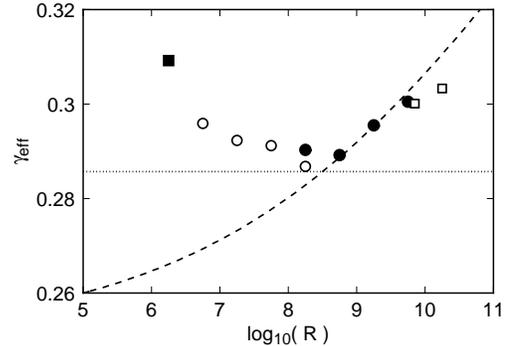}}
\caption{The effective exponents as a function of $R$ as in Fig.~\ref{fig:gamma}, but based on using ${\cal N} -1 $ in the analysis rather than $\cal N$. The symbols are as in Fig.~\ref{fig:gamma}. The dashed line is the logarithmic derivative of Eq.~\ref{Eq:powerlaw2} with $a = 0.311$ and $b = 2.58\times 10^{-3}$, which corresponds to the dashed line in Fig.~\ref{fig:GLplot}.}
\label{fig:gamma2} 
\end{figure}

Local fits of a powerlaw to data for ${\cal N} - 1$ are shown in Fig.~\ref{fig:gamma2}. There we also show, as a dashed line,  $\gamma_{eff}$ based on the dashed line in Fig.~\ref{fig:GLplot}. We see that deviations from the prediction Eq.~\ref{Eq:powerlaw2} already occur near $R = 10^8$, which is an order of magnitude larger than was found in the analysis using $\cal N$. However, for $R \agt 10^8$ the fit of the theory to the data is equally good, regardless of whether $\cal N$ or ${\cal N}-1$ is used.
\narrowtext

\begin{figure}
\epsfxsize = 2.75in
\centerline{\epsffile{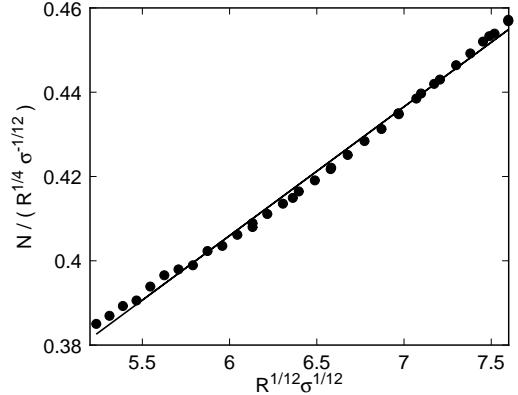}}
\caption{Plot of ${\cal N}/(R^{1/4}\sigma^{-1/12})$ as a function of $R^{1/12}\sigma^{1/12}$. If Eq.~\ref{Eq:powerlaw3} is correct, the points should fall on the straight line, a least-squares fit. The data deviate systematically, showing that Eq.~\ref{Eq:powerlaw3} is not the right functional form.}
\label{fig:GLplot2} 
\end{figure}

For large $\sigma$, the theory predicts crossover from $I_u$ to $III_u$ and   Eq.~\ref{Eq:powerlaw2}.\cite{GL00}  For somewhat smaller $\sigma \simeq 1$, however, the crossover should be from $I_u$ to $IV_u$, and the prediction then reads\cite{GL00} 
\begin{equation}
{\cal N} = a \sigma^{-1/12} R^{1/4} + b R^{1/3} \ \ .
\label{Eq:powerlaw3}
\end{equation}
As a function of $\sigma$, GL estimate that the transition from Eq.~\ref{Eq:powerlaw2} to Eq.~\ref{Eq:powerlaw3} occurs near $\sigma = 2$, but there is some uncertainty in this value. Thus, in Fig.~\ref{fig:GLplot2} we compared Eq.~\ref{Eq:powerlaw3} with our results for $\Gamma = 1$ by plotting ${y = \cal N}/(R^{1/4}\sigma^{-1/12})$ as a function of $x = R^{1/12}\sigma^{1/12}$. If Eq.~\ref{Eq:powerlaw3} is applicable, this should yield a straight line. As can be seen, the data deviate systematically from the fit. Thus a transition from $I_u$ to $IV_u$ at $\sigma = 4$ is inconsistent with our data.
\narrowtext
\begin{figure}
\epsfxsize = 3.0in
\centerline{\epsffile{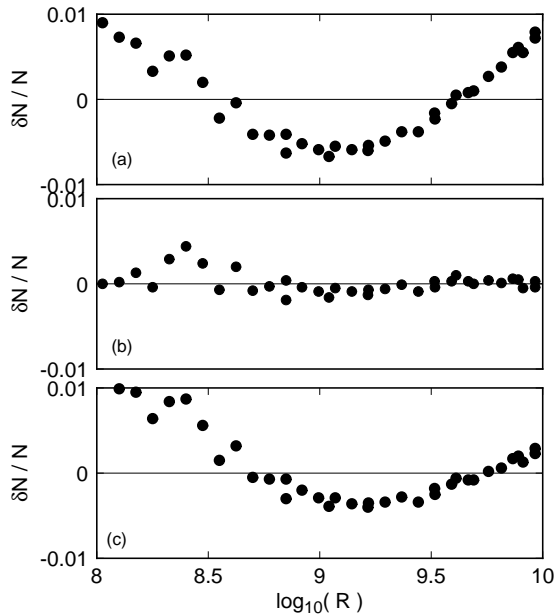}}
\caption{Relative deviations $\delta{\cal N}/{\cal N}$ from fits of (a): Eq.~\ref{Eq:powerlaw1}, (b): Eq.~\ref{Eq:powerlaw2}, and (c): Eq.~\ref{Eq:powerlaw3} to the $\Gamma = 1.0$ data.}
\label{fig:deviations} 
\end{figure}

In Fig.~\ref{fig:deviations} we show the deviations $\delta {\cal N}/{\cal N}$ from fits of Eqs.~\ref{Eq:powerlaw1} to Eq.~\ref{Eq:powerlaw3} to the $\Gamma = 1.0$ data. For Eq.~\ref{Eq:powerlaw1}  they are systematic, as already expected. For Eq.~\ref{Eq:powerlaw2} the fit is nearly perfect. We conclude that our data are consistent with Eq.~\ref{Eq:powerlaw2}. For Eq.~\ref{Eq:powerlaw3} the deviations are nearly as large as those for Eq.~\ref{Eq:powerlaw1}. Thus Eq.~\ref{Eq:powerlaw3} is not applicable to the data. A similar analysis of the $\Gamma = 0.5$ data over the range $10^9 \alt R \alt 4\times10^{10}$ gives similar results.

So far our work has concentrated on measurements of ${\cal N}(R)$ for acetone with $\sigma = 4.0$. We concluded  that there is no significant range of $R$ over which the  powerlaw Eq.~\ref{Eq:powerlaw1} is applicable, and that the crossover function Eq.~\ref{Eq:powerlaw2} proposed by Grossmann and Lohse provides a good fit to the data for $R \agt 10^7$ where the Reynolds number $R_e$ of the large-scale flow is expected\cite{GL00} to exceed about 50. Obviously a great deal of additional high-precision work remains to be done, and some of it is indeed under way. To provide further tests of the GL predictions, we started a systematic study of the Nusselt number ${\cal N}(R,\sigma)$ as a function of $\sigma$, but this work encountered considerable obstacles. For acetone $\sigma$ does not vary enough with $T$, and systematic errors in the properties\cite{properties} of other suitable fluids such as the alcohols prevent a comparison before better measurements of $\lambda$ and $\alpha/\kappa \nu$ are made. Since $R_e$ is central to the GL theory, its determination is also an obvious area for further work. And finally, measurements of comparable accuracy and precision should be extended to the compressed gases where $\sigma \simeq 0.7$, where more direct comparison with much previous work\cite{Si94,FN1} is possible, and where a different GL crossover function\cite{GL00} should pertain. 

We are grateful to Siegfried Grossmann and Detlef Lohse 
for stimulating correspondence, and to David Cannell for assistance in establishing our temperature scale. 
This work was supported by the 
National
Science Foundation through grant DMR94-19168.

\end{multicols}

\begin{references}

\bibitem{HCL87}F. Heslot, B. Castaing, and A. Libchaber, Phys. Rev. A {\bf 36}, 5870 (1987).

\bibitem{CGHKLTWZZ89}B. Castaing, G. Gunaratne, F. Heslot, L. Kadanoff, A. Libchaber, S. Thomae, X.Z. Wu, S. Zaleski, and G. Zanetti, J. Fluid Mech. {\bf 204}, 1 (1989). 

\bibitem{Si94}For reviews, see E.D. Siggia, Annu. Rev. Fluid Mech. {\bf 26}, 137 (1994); and S. Zaleski, in {\it Geophysical and Astrophysical Convection}, edited by P. Fox and R. Kerr (Gordon and Breach, New York, 1998).

\bibitem{FN} Data taken with helium gas by Chavanne et al. [X. Chavanne, F. Chilla, B. Chabaud, B. Castaing, J. Chaussy, and B. H\'ebral, J. Low Temp. Phys. {\bf 104}, 109 (1996)] show a departure from a powerlaw with $\gamma = 2/7$ for $R > 3\times10^{10}$. However, these results differ from those of Niemela {\it et al.} \cite{NSSD00}, which were also obtained with helium gas. The results of Cioni {\it et al.} [S. Cioni, S. Ciliberto, and J. Sommeria, J. Fluid Mech. {\it 335}, 111 (1997)] for liquid mercury also suggest a deviation from Eq.~\ref{Eq:powerlaw1}, but these results are for the very different Prandtl number regime $\sigma \simeq 0.025$. 

\bibitem{NSSD00}J.J. Niemela, L. Skrbek, K.R. Sreenivasan, and R.J. Donnelly, to be published.

\bibitem{FN1}A detailed examination of the recent experimental and theoretical literature was presented recently in Ref. \cite{GL00}.

\bibitem{GL00}S. Grossmann and D. Lohse, J. Fluid Mech. {\bf 400}, in print.

\bibitem{SS90}B.I. Shraiman and E.D. Siggia, Phys. Rev. A {\bf 42}, 3650 (1990).

\bibitem{DAC95}M.A. Dominguez-Lerma, G. Ahlers, and D.S. Cannell, Phys. Rev. E. {\bf 52}, 6159 (1995).

\bibitem{properties}T.E. Daubert and R.P. Danner, {\it Physical and Thermodynamic Properties of Pure Chemicals} (Hemisphere Publishing Corporation, New York, 1989).

\bibitem{BA00}K.M.S. Bajaj and G. Ahlers, unpublished.

\bibitem{WL91}X.-Z. Wu and A. Libchaber, Phys. Rev. A {\bf 43}, 2833 (1991).

\bibitem{Bu67}F. Busse, J. Fluid Mech. {\bf 30}, 625 (1967).

\bibitem{WL92}X.-Z. Wu and A. Libchaber, Phys. Rev. A {\bf 45}, 842 (1992).

\bibitem{Cetal}X. Chavanne {\it et al.}, in Ref. \cite{FN}.

\bibitem{AS00}S. Ashkenazi and V. Steinberg, Phys. Rev. Lett. {\bf 83}, 3641 (1999).

\bibitem{LE97}Y. Liu and R.E. Ecke, Phys. Rev. Lett. {\bf 79}, 2257 (1997).

\bibitem{Lohse} We are grateful to Detlef Lohse for suggesting this graphical representation.

\end{references}
\end{document}